\documentclass[11pt]{article}
\usepackage[margin=1in]{geometry}
\usepackage{amsmath,amssymb}
\usepackage{graphicx}
\usepackage{booktabs}
\usepackage{hyperref}
\usepackage{natbib}
\usepackage{xcolor}
\usepackage{listings}

\title{Symbolic Reasoning Frameworks Trigger Memory-Mediated\\
\large Ecosystem Dynamics in Multi-Agent LLM Systems}
\author{Augustin Chan\\
\texttt{aug@iterative.day}}
\date{June 2026 (v2)}

\begin{document}
\maketitle

\begin{abstract}
Large language models exhibit innate behavioral tendencies when deployed as strategic agents---notably a risk-averse ``turtle'' bias toward defensive play. We show that injecting a symbolic reasoning framework as a per-round reflective prompt into a single agent acts as a small cognitive perturbation whose consequences are not per-decision but \emph{emergent}: the receiving agent's risk posture is unchanged in isolation, yet over a full campaign of accumulating memory and multi-agent interaction the conditions settle into different, condition-associated winner ecosystems---an effect we attribute to emergent memory and multi-agent dynamics rather than the framework's per-decision influence. In a 7-player Warring States Diplomacy variant (61 games, 6 conditions, single-campaign memory accumulation), the winner distribution differs sharply across the four primary conditions (41 games; Monte-Carlo permutation omnibus $p \approx 0.001$), forming condition-associated ecosystem signatures: under control, Yan dominates (7/11, 64\%); under I-Ching yarrow divination, Yan and Chu co-dominate while Qin is completely suppressed (0/10); under Tarot, Qin dominates (5/10); under scrambled-text ablation (English commentary word-shuffled; hexagram name and Chinese judgment retained), Qi dominates (5/10). The scrambled$\to$Qi attractor is the most robust (significant against both a pooled denominator and control alone, $p = 0.006$ and $0.012$); the tarot$\to$Qin attractor is significant only against the pooled denominator ($p = 0.006$, vs.\ control alone $p = 0.064$). The framework-receiving agent (Han) never wins and shows no survival difference across conditions (Fisher $p = 1.0$), but Tarot consistently elevates Han's peak territory (mean 3.0 SCs vs.\ 2.1--2.5 others, Kruskal-Wallis $p = 0.010$). Neither framework's content predicts subsequent actions---hexagram themes ($\chi^2$ $p = 0.95$) and Tarot card postures ($\chi^2$ $p = 0.69$) are both independent of action choice. A memory-free decision-isolation probe (960 calls) further shows the modulation is \emph{emergent rather than per-decision}: stripped of memory and multi-agent context the reflective process does not change the receiving agent's risk posture (hold-rate Friedman $p = 0.45$; the I-Ching changes no decisions beyond temperature noise, $p = 0.60$; Tarot perturbs move content but not risk posture, $p = 0.021$), locating the effect in memory accumulation and multi-agent dynamics rather than per-decision processing. A 2$\times$2 factorial decomposition separating yarrow's decision-time (per-round oracle) and learning-time (between-game I-Ching reflection) components reveals a non-additive interaction: each component individually freezes the board (50--60\% stalemate rate, Fisher $p = 0.012$ and $0.004$ vs.\ control), but combined they produce zero stalemates (negative interaction, permutation $p \approx 5\times10^{-5}$). Direct testing relocates yarrow's Qin suppression from the framework-receiving agent's own cooperation to rival (Chu) expansion through a map corridor; a depth-matched follow-up then shows that expansion is governed by campaign memory depth, not the oracle (condition $p = 0.55$), so even this pathway is emergent rather than framework-specific. We present this as an observation paper establishing that alignment-framework choice at the agent level produces distinctive, non-additive system-level consequences in multi-agent settings, transmitted through emergent multi-agent and memory dynamics rather than the framework's per-decision effect on the receiving agent.
\end{abstract}

\section{Introduction}
\label{sec:intro}

Large language models deployed as strategic agents exhibit innate behavioral tendencies. Recent work documents these biases across multiple game settings: LLMs split into behavioral archetypes in board games \citep{Jain2026ludobench}, display strong fairness preferences in dictator games \citep{Einwiller2025dictator}, and exhibit model-specific strategic profiles in civilization-building \citep{Chen2026civbench}. These are not bugs---they are default behavioral modes that emerge from training.

A natural question follows: can these tendencies be modulated? The persona-prompting literature answers yes---assigning traits (``be aggressive'') or steering activations shifts strategic behavior \citep{Licato2025persona, Sun2026persona}. But this literature measures effects on the \emph{prompted agent}. In multi-agent settings, the more consequential question is: does modulating one agent's behavior change \emph{other agents' outcomes}?

We investigate this using symbolic reasoning frameworks---not trait labels or behavioral instructions, but philosophical systems requiring interpretation. Specifically, we compare the I-Ching (ancient Chinese divination via yarrow-stalk casting, producing hexagrams with abstract commentaries) and Tarot (three-card spreads with situational interpretations) as per-round reflective prompts for one agent in a 7-player strategic game. A scrambled-text ablation (the I-Ching commentary word-shuffled, with the hexagram name and Chinese judgment retained---a coherence-degraded, not content-free, oracle) and a generic-reflection control complete the design.

The distinction between frameworks and personas is load-bearing. A persona says ``be cooperative.'' A framework says ``Hexagram 36, Brightness Hiding: the light retreats into the earth; the wise person dims their brilliance among the masses'' and asks the agent to interpret this in context. The interpretation step is where we initially located the mechanism---but, as \S\ref{sec:content} shows, the effect turns out to be \emph{emergent} rather than a per-decision property of that interpretation.

Our core empirical finding is system-level: the symbolic framework injected into one agent is only a small perturbation, but over a campaign of accumulating memory and multi-agent interaction it is associated with a distinctive, condition-dependent reorganization of the \emph{entire} ecosystem. The fingerprint is in the winner distribution. Under I-Ching yarrow divination, Qin (the strongest expansionist) wins 0 of 10 games while Yan and Chu co-dominate. Under Tarot, Qin wins 5 of 10 (Fisher $p = 0.006$ vs.\ pooled). Under scrambled text, Qi wins 5 of 10 ($p = 0.006$). Under control, Yan dominates with 7 of 11 wins. The framework-receiving agent (Han) never wins under any condition---the conditions differ in ecosystem outcomes, not in Han's fate.

The modulation is neither content-following nor a per-decision process effect: hexagram themes do not predict the agent's subsequent actions ($\chi^2$ $p = 0.95$) even though 77\% of orders reference the hexagram, and a memory-free decision-isolation probe (\S\ref{sec:content}) shows the reflective process does not modulate the receiving agent's risk posture in isolation. The scrambled-text ablation further argues against prompt structure alone. The effects are thus \emph{emergent}---properties of memory accumulation and the multi-agent interaction more than of coherent symbolic content per se (scrambled retains a coherent name and Chinese judgment, so it bounds rather than eliminates content effects; \S\ref{sec:limitations}). On this reading the symbolic framework is the perturbation and the ecosystem reorganization is the phenomenon, with the amplification supplied by emergent memory and multi-agent dynamics; we do not isolate which of those two channels carries the effect. The picture parallels---and, we suggest, extends---2026 findings that accumulated history dominates per-turn prompting in multi-agent LLM systems \citep{MemoryCurse2026}: if memory dominates prompting, the small differences a reasoning framework introduces early could accumulate into divergent ecosystem-level attractors. The single-agent claim ``memory changes the agent'' would become a collective one---memory changing the entire ecology. We present this as the reading our results point toward rather than an isolated mechanism.

This paper is a companion to a negative-result study of the King Wen I-Ching sequence in neural network training \citep{Chan2026kingwen}. There, the sequence's anti-habituation properties failed to improve training. Here, the same sequence (via yarrow-stalk casting) fails to improve strategic decision-making but produces measurable ecosystem effects. The sequence appears to be a reliable perturbator---it changes behavior without improving it.

\section{Related Work}
\label{sec:related}

\paragraph{LLM behavioral biases in strategic games.}
LLMs exhibit systematic behavioral tendencies in game-theoretic settings. LudoBench finds LLMs agree with game-theory baselines only 40--46\% of the time \citep{Jain2026ludobench}. LLMs display strong fairness preferences in dictator games \citep{Einwiller2025dictator}. GPT-3.5 can operationalize cooperative vs.\ competitive personas with conditional reciprocity \citep{Phelps2023psychology}. Llama reproduces population-level human cooperation patterns while Qwen aligns with Nash equilibrium---different models exhibit fundamentally different innate tendencies \citep{CeraPalatsi2025cooperation}. Bias-adjusted agents using behavioral economics frameworks shift decision-making toward human-like patterns \citep{Kitadai2025bias}. CivBench finds model-specific strategic profiles not visible through outcome-only evaluation \citep{Chen2026civbench}.

\paragraph{Persona and prompt effects on strategic behavior.}
Direct persona assignment does not reliably transfer to strategy without a mediator \citep{Licato2025persona}. Activation-vector steering shifts both strategy and justification, with rhetoric and strategy sometimes diverging \citep{Sun2026persona}---directly relevant to our Tarot result (reasoning elevated without outcome improvement). Assigning human-like identity does not produce human-like behavior \citep{Ma2024behavioral}. Role-playing agents show systematic inconsistencies between stated beliefs and simulated behavior \citep{Mannekote2025roleplaying}.

\paragraph{Multi-agent LLM strategic systems.}
LLMs exhibit high baseline behavioral similarity in strategic settings (strategic algorithmic monoculture; \citealt{Ballestero2026monoculture}). In Diplomacy, combining language models with strategic reasoning achieves human-level play \citep{Bakhtin2022diplomacy}, self-evolving agents with memory reach competent play \citep{Guan2024richelieu}, fine-tuned LLMs learn equilibrium policies \citep{Xu2025dipllm}, and fine-grained analysis reveals LLMs and humans differ systematically in negotiation tactics \citep{Li2025negotiation}. SAE analysis discovers reward-hacking behaviors in Diplomacy training \citep{Yan2026interpretability}. Coalition formation analysis shows emergent stability properties in LLM agent networks \citep{Guo2026coalition}.

\paragraph{Memory and history as dominant drivers.}
A 2026 line of work establishes that in repeated multi-agent settings, accumulated interaction history dominates per-turn prompting. \citet{MemoryCurse2026} show across seven models and four social dilemmas (378k traces) that expanding historical memory degrades cooperation---agents become ``risk-minimizing and history-following''---and, via sanitization experiments holding prompt length fixed, that \emph{memory content dominates over prompt and persona}; notably, 10 of 28 model-game settings are ``memory-immune,'' a strong heterogeneity. History \emph{structure} is itself a design lever, with non-monotonic histories amplifying path dependence \citep{NetworkHistory2025}, and recent surveys position memory as what turns a stateless generator into an adaptive agent \citep{MemorySurvey2026}. Our results are the cross-agent, ecosystem-level analog: the framework-receiving agent's per-decision behavior is not modulated by the reflective process in isolation (\S\ref{sec:content}), and a key board-level outcome is governed by campaign memory depth rather than the oracle (\S\ref{sec:mechanisms}).

\paragraph{Risk aversion and strategic generalization.}
Risk-averse agents exhibit less free-riding and better equilibrium outcomes with unseen partners \citep{Qu2026riskaversion}, providing theoretical grounding for our finding that differential modulation of risk aversion produces distinct ecosystem-level outcomes.

\paragraph{The gap.}
No published work uses a philosophical/symbolic reasoning framework---requiring \emph{interpretation} rather than instruction-following---as the modulating mechanism. No work measures ecosystem-level winner distributions as a function of one agent's reasoning framework. And the distinction between modulation via framework \emph{content} and framework \emph{process} has not been tested.

\section{Methods}
\label{sec:methods}

\subsection{Game Environment}

The game is a 7-player variant of Diplomacy played on a 46-territory map based on the Chinese Warring States period (475--221 BCE). The map includes 27 supply centers (19 state home SCs + 8 neutral) and 19 non-SC corridor territories. Each state begins with one unit per home supply center (2--4 units, depending on historical territory). Victory requires 14 of the 27 supply centers (``domination'') or the most supply centers after 20 rounds; ties are broken by army count, and an exact tie is recorded as a draw (one scrambled-condition game ran a single round past the limit, terminating at round 21). Combat uses Diplomacy-style mechanics---simultaneous hidden orders, support orders, and deterministic resolution---with two simplifications relative to standard Diplomacy: dislodged units retreat to the first available province, and each state makes at most one build or disband per round.

\subsection{Agents}

Each of the 7 states is controlled by an independent Claude Opus 4.6 agent instance with a state-specific persona prompt grounded in historical and philosophical sources (\emph{Shiji}, \emph{Han Feizi}, \emph{Zhanguoce}). Table~\ref{tab:states} summarizes the seven states.

\begin{table}[t]
\centering
\caption{Seven states and their LLM persona characteristics. Han$^*$ is the intervention recipient.}
\label{tab:states}
\begin{tabular}{llll}
\toprule
State & School & Advantage & Disadvantage \\
\midrule
Qin & Legalism & Attack, reform & Alliance trust \\
Han$^*$ & King Wen & Adaptability & Smallest, weakest \\
Wei & Administration & Early game, talent & Exposed geography \\
Zhao & Militarism & Cavalry, defense & Stability \\
Qi & Eclecticism & Intelligence, income & Recovery after losses \\
Chu & Daoism & Depth, territory & Reform efficiency \\
Yan & Confucianism & Defense, surprise & Economy, passivity \\
\bottomrule
\end{tabular}
\end{table}

\subsection{Intervention Design}
\label{sec:intervention}

The intervention operates at two timescales:

\paragraph{Decision-time (per-round).} Before each order submission, Han receives oracle text together with a \textsc{MANDATE} instruction to interpret the oracle in relation to the current strategic situation before issuing orders. The control condition receives a length-matched generic reflection prompt without oracle content.

\paragraph{Learning-time (between-game).} After each game, Han reflects on the game through the lens of its assigned framework. Reflections are stored in a SQLite memory bank and retrieved in subsequent games within the same campaign.

\paragraph{Hexagram text variation.} Due to a corpus configuration change mid-campaign, the first 6 yarrow games provided only hexagram numbers and line diagrams (name listed as ``Unknown,'' judgment and commentary fields empty), while the last 4 provided the full hexagram name, Chinese source text, and English commentary. In the number-only games, Claude supplied hexagram names and meanings from its training data (e.g., producing ``Hexagram 18 (Gu/Decay)'' from the input ``Hexagram 18, Unknown''). The ecosystem signature---Qin suppression (0/6 and 0/4), Yan/Chu co-dominance---held across both subgroups, providing an unplanned within-condition ablation of oracle \emph{content} that further supports the content-independence finding (\S\ref{sec:content}).

\subsection{Memory Architecture}
\label{sec:memory}

Because the headline effects are properties of accumulating memory (\S\ref{sec:content}, \S\ref{sec:discussion}), and memory now figures in our central claim, we describe the implementation explicitly rather than leaving it to the appendix example. Memory is a per-agent SQLite store of structured \emph{insights}; each record is a (\emph{situation}, \emph{lesson}, confidence~$\in\{$low, medium, high$\}$) triple with coarse situation features (territory bucket, threat level, position type) and, for treatment Han only, the hexagram context that framed the reflection.

\paragraph{Isolation.} Every memory is keyed by (campaign, agent, condition) and retrieved only for the matching key. All seven agents---not only Han---accumulate their own separate banks; no memory crosses agents, conditions, or campaigns. The framework asymmetry is thus in \emph{content} (only Han's oracle prompt and hexagram-framed reflection differ), not in the memory \emph{mechanism}, which is identical for every agent.

\paragraph{Write.} After each completed game, every agent runs a between-game reflection (Claude Opus 4.6) over a compressed timeline and summary of that game and returns structured insights, appended to its bank. The bank therefore grows with campaign position---the memory-depth axis the breach (\S\ref{sec:mechanisms}) and decision-isolation (\S\ref{sec:content}) analyses turn on.

\paragraph{Read.} Before issuing orders, an agent is injected with up to $K=8$ of its active memories, selected \emph{deterministically} by a weighted score---relevance~$0.5$ $+$ recency~$0.3$ $+$ confidence~$0.2$, where relevance is Jaccard overlap of situation features---capped at two memories per source game for diversity and formatted as a ``lessons from previous games'' block with citation tags (these tags enable the citation-rate and content-independence analyses of \S\ref{sec:content}). Only the reflection text itself is model-generated; given a fixed bank and board, retrieval is deterministic.

\paragraph{Curation and reset.} Insights carry usage counters and an \emph{active} flag; a newer insight can supersede an older one (deactivating it), so the eligible bank is curated rather than monotonically growing. A new campaign uses a fresh store---memory never crosses campaigns. This is what makes each condition a single accumulating history, and what the memory-free decision-isolation probe (\S\ref{sec:content}) removes by construction.

\subsection{Conditions}
\label{sec:conditions}

Six conditions, each modifying only Han's prompt (Table~\ref{tab:conditions}): four primary conditions plus two factorial-decomposition conditions (\S\ref{sec:factorial-design}) that separate the yarrow intervention's decision-time and learning-time components.

\begin{table}[h]
\centering
\caption{Experimental conditions. All conditions modify only Han's prompt; the other six agents are identical across conditions. The first four rows are the primary conditions ($n=41$) carrying the ecosystem-signature analysis (\S\ref{sec:ecosystem}); decision-only and learning-only are the factorial-decomposition conditions (\S\ref{sec:factorial-design}) that isolate the per-round and between-game components of the yarrow intervention---61 games total.}
\label{tab:conditions}
\begin{tabular}{lp{7.0cm}l}
\toprule
Condition & Han's Per-Round / Between-Game Prompt & $n$ \\
\midrule
Control & Length-matched generic reflection prompt & 11 \\
Yarrow & Yarrow-stalk I-Ching cast; hexagram text + \textsc{MANDATE} to interpret & 10 \\
Tarot & 3-card Tarot spread (situation / hidden influence / posture) + \textsc{MANDATE} & 10 \\
Scrambled & Yarrow structure; English commentary word-shuffled (name + Chinese judgment intact) + \textsc{MANDATE} & 10 \\
Decision-only & Yarrow \textsc{MANDATE} per round; \emph{generic} reflection between games & 10 \\
Learning-only & \emph{Generic} prompt per round; I-Ching framework reflection between games & 10 \\
\bottomrule
\end{tabular}
\end{table}

The scrambled condition was \emph{designed} as a content-coherence ablation: preserve the structural intervention (oracle-formatted prompt, \textsc{MANDATE} to interpret) and the hexagram number and name (including the Chinese name) as structural identifiers, while word-shuffling the judgment and commentary to destroy coherent content. In practice it is \emph{degraded}, not content-free, for two reasons we flag honestly (Appendix~\ref{app:scrambled}). First, the shuffle is \emph{whitespace-based}, so it is a no-op on Classical Chinese judgment text, which has no inter-word spaces: for hexagrams whose source judgment is Chinese, a coherent Chinese judgment survived unshuffled---an implementation artifact, since the design intended to garble it. Second, the coherent English name (e.g.\ ``Peace'') is always present by design. So the agent---which reads Classical Chinese---received a valenced label and, for many casts, a coherent Chinese judgment, with only the English commentary reliably garbled. The intended contrast was: if effects derive from coherent oracle \emph{content}, scrambled should resemble control; if from \emph{structure}, scrambled should resemble yarrow. Neither holds---scrambled produces its own distinct ecosystem signature---but because coherent content survived, scrambled \emph{bounds} rather than eliminates the role of content (\S\ref{sec:limitations}).

\subsection{Factorial Decomposition Conditions}
\label{sec:factorial-design}

The yarrow intervention operates at two timescales (\S\ref{sec:intervention}): a per-round oracle (decision-time) and I-Ching-framed reflection between games (learning-time). To isolate them we add two conditions, yielding a 2$\times$2 factorial with control and full yarrow as the remaining cells (Table~\ref{tab:factorial-design}). \textbf{Decision-only} gives Han the yarrow \textsc{MANDATE} prompt per round but \emph{generic} between-game reflection; \textbf{learning-only} gives Han a \emph{generic} per-round prompt (identical to control) but I-Ching-framework reflection between games. This tests whether the full yarrow ecosystem signature is carried by the per-round oracle, the learning framework, or their interaction.

\begin{table}[h]
\centering
\caption{2$\times$2 factorial decomposition of the yarrow intervention.}
\label{tab:factorial-design}
\begin{tabular}{lcc}
\toprule
 & No oracle per-round & Oracle per-round \\
\midrule
No framework reflection & Control & Decision-only \\
Framework reflection & Learning-only & Yarrow (full) \\
\bottomrule
\end{tabular}
\end{table}

\subsection{Dataset}

61 games across 6 conditions: 11 control, 10 each yarrow, Tarot, scrambled text, decision-only, and learning-only. The four primary conditions (control, yarrow, Tarot, scrambled; $n=41$) carry the ecosystem-signature analysis; the two factorial conditions extend it to the decomposition in \S\ref{sec:factorial}. All agents use Claude Opus 4.6. Each condition runs as a single campaign with continuous memory accumulation across games (SQLite memory bank carries learning forward).

\subsection{Statistical Methodology}

Winner distributions: we first establish global heterogeneity with a Monte-Carlo permutation omnibus (100{,}000 reshuffles of the condition labels across all 41 games, evaluated under both a Pearson chi-squared and a summed pairwise total-variation statistic), since the condition$\times$winner table has numerous zero and small cells that invalidate the asymptotic chi-squared test. Pairwise winner comparisons then use Fisher's exact tests, reported against both a pooled ``vs.\ all other conditions'' denominator and the conservative direct ``vs.\ control'' denominator; pairwise condition separability is assessed by total-variation distance against the same permutation null. Behavioral rates: pooled proportions and per-game averages (game as unit of analysis) with Kruskal-Wallis and Mann-Whitney $U$ tests. Pairwise tests follow significant omnibus tests as planned contrasts with Bonferroni correction. Hexagram-action and Tarot-action association: Pearson's chi-squared and Fisher's exact. The factorial stalemate interaction is assessed by a label-permutation test on the difference-in-differences contrast across the four cells, since the empty cells make an asymptotic logistic (or Firth) interaction degenerate. We report exact $p$-values throughout.

Territorial statistics use final-game-state methodology: each game's last recorded standings.

\section{Results}
\label{sec:results}

\subsection{Behavioral Baseline: The LLM Turtle Tendency}
\label{sec:turtle}

Under the control condition, Claude Opus defaults to 44.3\% defensive play (hold orders) with 45.2\% move, 6.6\% self-support, and 3.9\% cooperative support ($n = 228$ orders across 11 games). This tendency intensifies over the course of a game: control Han's defensive rate rises from 40.5\% in the early game to 45.0\% in the late game, while move rate drops from 54.2\% to 37.5\%. We refer to this as the ``turtle tendency'': an innate bias toward holding position that strengthens under uncertainty, consistent with findings that RLHF-trained models exhibit passivity biases in competitive settings \citep{Mukobi2023welfare}.

\subsection{Framework-Specific Behavioral Modulation}
\label{sec:behavioral}

Each framework produces a qualitatively distinct behavioral profile in Han (Table~\ref{tab:behavior}).

\begin{table}[t]
\centering
\caption{Han behavioral profiles by condition. Hold rate = defensive orders; move = territorial repositioning/expansion; self-support = supporting own units; other-support = supporting another state's unit. Reasoning = mean characters per order in Han's strategic text.}
\label{tab:behavior}
\begin{tabular}{lcccc}
\toprule
& Control & Yarrow & Tarot & Scrambled \\
\midrule
Hold rate & 44.3\% & 53.7\% & \textbf{61.2\%} & 54.7\% \\
Move rate & 45.2\% & 33.2\% & \textbf{19.7\%} & 30.0\% \\
Self-support & 6.6\% & 7.9\% & \textbf{17.1\%} & 12.3\% \\
Other-support & 3.9\% & 5.1\% & 2.0\% & 2.9\% \\
Reasoning (chars) & 309 & 449 & 456 & \textbf{633} \\
\bottomrule
\end{tabular}
\end{table}

\paragraph{Yarrow: late-game cooperative pivot.} Yarrow-Han's aggregate profile differs modestly from control ($51.6\%$ vs.\ $43.2\%$ per-game hold rate, MWU $p = 0.18$). But in the late game, yarrow-Han increases cooperative support to 17.1\%---vs.\ 7.5\% control, 0.0\% Tarot, 3.6\% scrambled. Yarrow is the only condition under which late-game other-support reaches double digits.

\paragraph{Tarot: extreme risk aversion.} Tarot-Han holds or supports itself in 78.3\% of orders. Hold rate: control vs.\ Tarot MWU $p = 0.008$. Move rate: 4-way KW $p = 0.005$; control vs.\ Tarot $p = 0.0008$.

\paragraph{Scrambled: reactive self-reinforcement.} Scrambled-Han's late game is dominated by self-support (25.0\%, highest of any condition) and shows the most extreme pressure reactivity ($+12.3$pp defensive shift under territory loss, vs.\ $+1.8$pp yarrow, $+0.0$pp Tarot ceiling effect, $+8.9$pp control; Table~\ref{tab:pressure}).

\paragraph{Reasoning-length gradient.} Han's reasoning length follows a clear gradient: control 309, yarrow 449, tarot 456, scrambled 633 chars (4-way KW $p = 0.007$). Yet strategic outcomes do not improve with reasoning length---the more perturbative the prompt, the more deliberation it induces, but this deliberation does not translate into better outcomes.

\subsection{Content-Action Independence}
\label{sec:content}

A central concern is whether the behavioral modulation is simply content-following. We classify all 64 hexagrams by primary theme (advance/retreat/wait/cooperate) and test for association with Han's subsequent action category. The association is null: Pearson's $\chi^2$ $p = 0.9454$ ($n = 214$ orders, 4 themes $\times$ 4 actions, dof $= 9$). Fisher's exact for advance vs.\ non-advance hexagrams: OR $= 1.40$, $p = 0.34$. However, 77.1\% of Han's orders reference the hexagram in their reasoning text (131/170 yarrow orders with hexagram data)---Han reads and contemplates the hexagram but does not follow it as an instruction. This independence is especially striking for the 6 yarrow games where no hexagram text was provided: the agent generated its own interpretation from the hexagram number alone, then did not follow that self-generated interpretation either.

We perform an analogous test on the Tarot condition. Each card has a grounded \texttt{decision\_posture} (advance/hold/retreat/ally/transform/observe). The dominant posture of the 3-card spread does not predict Han's subsequent action: $\chi^2$ $p = 0.6860$ ($n = 299$ orders, 6 postures $\times$ 4 actions, dof $= 15$). Tarot-Han references card names in 81.6\% of reasoning text yet does not follow the cards' posture recommendations.

The scrambled-text ablation (\S\ref{sec:conditions}) degrades the commentary's coherence while preserving prompt structure (and the hexagram name and Chinese judgment): it produces its own distinct ecosystem signature (Qi dominance, 5/10, $p = 0.006$), arguing against \emph{coherent-commentary} content-following and against prompt structure alone---though, because the name and Chinese judgment survive, it bounds rather than rules out content effects (\S\ref{sec:limitations}). Content-action independence rules out content-following, but it does not by itself establish that the reflective \emph{process} modulates decisions; that inference rests on campaign-level differences, which carry memory and multi-agent confounds. We test it directly next.

\paragraph{Memory-free decision isolation: the process does not modulate risk posture.}
To isolate the per-decision effect of the reflective process from memory accumulation and multi-agent propagation, we ran a pre-registered fixed-scenario probe. For 40 real board positions sampled across game phase, the framework-receiving agent was queried under three reflective blocks over an \emph{identical} board with \emph{no memory} (control = length-matched generic; yarrow = I-Ching; tarot = 3-card spread), 8 replicates each (960 calls, same model pin). As orders vary at temperature, the signal is judged against a control-vs-control noise floor (0.294 mean Jaccard order-divergence). Order-divergence from control exceeds the floor for tarot (0.368, Wilcoxon $p = 0.021$) but not yarrow (0.329, $p = 0.60$): tarot perturbs \emph{which} move is chosen, the I-Ching does not move decisions beyond noise. This asymmetry is confounded with prompt \emph{directiveness}: the Tarot spread states an explicit recommended posture for each card (e.g.\ ``Posture: retreat,'' Listing~\ref{lst:tarot}) while the I-Ching supplies abstract hexagram text to interpret (Listing~\ref{lst:yarrow}), so a more directive prompt may perturb the chosen move regardless of symbolic content or model--oracle cultural match. The posture is still not \emph{followed} (the posture--action $\chi^2$ above, $p = 0.69$); a directive \emph{format}, however, can shift the decision distribution without being obeyed. We therefore read the tarot/I-Ching gap as suggestive and revisit it under the form-controlled congruence design of \S\ref{sec:future}. Critically, on the paper's own dependent variable---the turtle hold rate (\S\ref{sec:turtle})---the arms are statistically indistinguishable (scenario-level Friedman $p = 0.45$; yarrow vs.\ control $p = 0.10$; tarot vs.\ control $p = 0.73$), and the risk-aversion ordering tarot $\geq$ yarrow $\geq$ control is not observed. What survives memory-free is reasoning length: both frameworks elevate it ${\sim}33\%$ (control 1029 $\to$ yarrow 1351, tarot 1382 characters). Thus the frameworks change \emph{how much} the agent deliberates and, for tarot, \emph{which} move it picks---but not its risk posture, in isolation. The risk-aversion modulation and the winner-ecosystem effects do not originate in the per-round reflective act; they are emergent properties of the campaign (memory accumulation) and the seven-agent interaction (\S\ref{sec:discussion}).

\subsection{Ecosystem Signatures: Winner Distributions}
\label{sec:ecosystem}

\begin{figure}[t]
\centering
\includegraphics[width=0.85\linewidth]{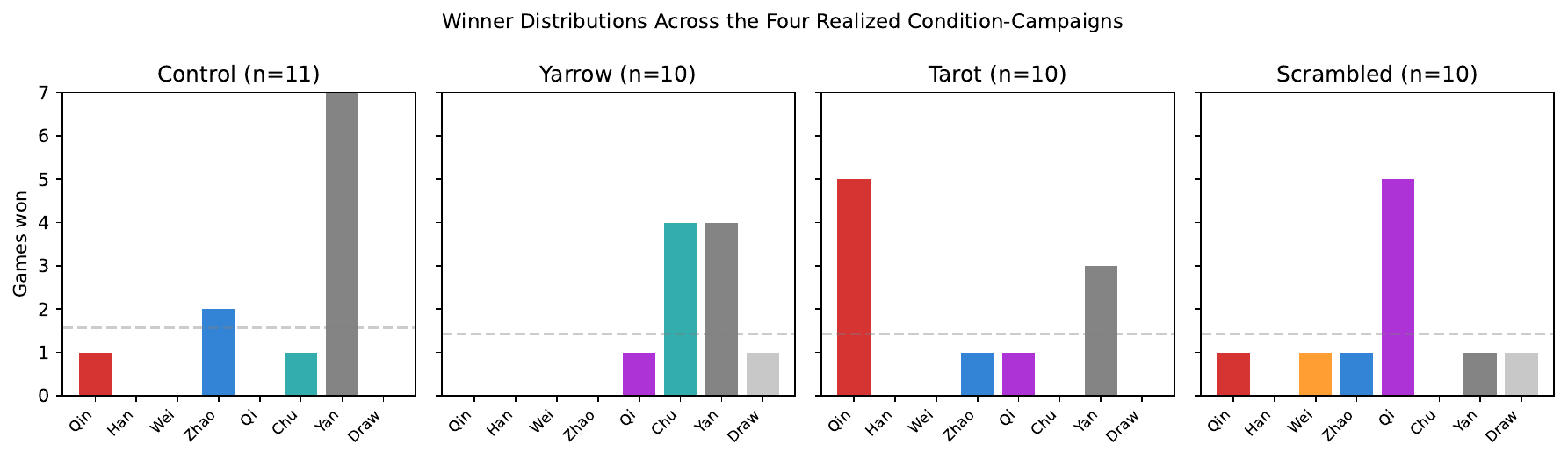}
\caption{Winner distributions across the four realized condition-campaigns. The global distribution is heterogeneous (permutation omnibus $p \approx 0.001$), though only the scrambled condition is individually separable at this sample size: control $\to$ Yan dominance, yarrow $\to$ Yan/Chu co-dominance with Qin suppressed, tarot $\to$ Qin dominance, scrambled $\to$ Qi dominance.}
\label{fig:winners}
\end{figure}

The behavioral differences propagate to other agents' outcomes, producing condition-associated winner-ecosystem signatures (Table~\ref{tab:winners}, Figure~\ref{fig:winners}). A Monte-Carlo permutation omnibus (100{,}000 reshuffles of the condition labels over the 41 games) rejects a common winner distribution at $p \approx 0.0013$ under both a chi-squared and a total-variation statistic, establishing strong within-sample heterogeneity across the four realized condition-campaigns before any pairwise claim. Because each primary condition was run as a single continuous campaign, this heterogeneity reflects each condition \emph{as run}---the intervention together with its one accumulated campaign---and does not by itself separate the framework's contribution from between-campaign variance, which is substantial (\S\ref{sec:mechanisms}); the controlled decision-isolation (\S\ref{sec:content}) and factorial (\S\ref{sec:factorial}) analyses bear on that separation.

\begin{table}[t]
\centering
\caption{Winner distributions by condition. Han never wins under any condition. Bold indicates the condition's dominant winner. Draws included in denominators.}
\label{tab:winners}
\begin{tabular}{lcccccccc}
\toprule
& Qin & Yan & Qi & Chu & Zhao & Wei & Draw & Han \\
\midrule
Control ($n\!=\!11$) & 1 & \textbf{7} & 0 & 1 & 2 & 0 & 0 & 0 \\
Yarrow ($n\!=\!10$) & \textbf{0} & 4 & 1 & \textbf{4} & 0 & 0 & 1 & 0 \\
Tarot ($n\!=\!10$) & \textbf{5} & 3 & 1 & 0 & 1 & 0 & 0 & 0 \\
Scrambled ($n\!=\!10$) & 1 & 1 & \textbf{5} & 0 & 1 & 1 & 1 & 0 \\
\bottomrule
\end{tabular}
\end{table}

The four condition-associated modal signatures are:
\begin{itemize}
\item \textbf{Control $\to$ Yan dominance} (7/11, 64\%): under no-framework conditions with full memory accumulation, Yan's defensive posture wins overwhelmingly.
\item \textbf{Yarrow $\to$ Yan/Chu co-dominance, Qin suppressed} (Yan 4, Chu 4, Qin 0): the I-Ching framework produces a cooperative/defensive ecosystem that shuts out the strongest expansionist.
\item \textbf{Tarot $\to$ Qin dominance} (5/10, 50\%): the Tarot framework produces conditions favorable to the strongest expansionist state.
\item \textbf{Scrambled $\to$ Qi dominance} (5/10, 50\%): the coherence-degraded oracle (English commentary shuffled; name and Chinese judgment intact) produces its own distinct ecosystem favoring the intelligence/income-advantaged state.
\end{itemize}

\paragraph{Pairwise separability is weaker than the global heterogeneity.} Decomposing the omnibus by total-variation distance against the same permutation null, only the scrambled condition separates from its neighbours at the uncorrected 5\% level (vs.\ control TV $= 0.71$, $p = 0.016$; vs.\ yarrow TV $= 0.70$, $p = 0.029$). The control--yarrow--tarot cluster is \emph{not} pairwise-distinguishable on winner identity alone at this sample size (every pair $p > 0.11$). We therefore frame the result as a single strongly heterogeneous landscape in which scrambled is the most distinct individual signature, rather than four mutually separable signatures.

\paragraph{The two attractor claims.} The tarot$\to$Qin and scrambled$\to$Qi effects were tested as pre-registered attractor hypotheses against a pooled ``all-other-conditions'' denominator, each $p = 0.006$ (5/10 vs.\ 2/31), surviving Bonferroni correction for the $m = 2$ family. Pooling, however, presumes the comparator conditions are exchangeable on the target state's win rate---an assumption the omnibus shows is only approximate---so we also report the conservative direct comparison against control. Under that test the two effects diverge: \textbf{scrambled$\to$Qi is robust} (significant against both the pooled denominator and control alone: 5/10 vs.\ 0/11, $p = 0.012$), whereas \textbf{tarot$\to$Qin is denominator-dependent} (5/10 vs.\ control 1/11, $p = 0.064$, not reaching $\alpha = 0.05$). A leave-one-out analysis shows the pooled tarot$\to$Qin result is insensitive to any single game ($p \in [0.003, 0.016]$, significant in all ten deletions); its fragility lies in the choice of denominator, not sample composition. Given the finding's replication history (\S\ref{sec:limitations}), we present tarot$\to$Qin as suggestive and in need of out-of-sample replication, and treat scrambled$\to$Qi as the stronger attractor.

Additional pairwise tests: yarrow Qin suppression (0/10) vs.\ Tarot (5/10): Fisher $p = 0.033$; yarrow Qin suppression vs.\ all others (7/31): Fisher $p = 0.161$.

\subsection{Ecosystem Mechanisms}
\label{sec:mechanisms}

How do Han's behavioral profiles propagate to other agents' outcomes? We analyze late-game support orders and pressure responses across conditions.

\paragraph{Late-game cooperation as friction.} Yarrow-Han's late-game other-support rate (17.1\%) is the clear outlier (Table~\ref{tab:latesupport}). Tarot-Han produces zero late-game cooperation across 10 games. Scrambled-Han's late game is dominated by self-support (25.0\%).

\begin{table}[t]
\centering
\caption{Late-game support behavior by condition.}
\label{tab:latesupport}
\begin{tabular}{lcccc}
\toprule
& Control & Yarrow & Tarot & Scrambled \\
\midrule
Late self-support & 10.0\% & 7.3\% & 13.6\% & \textbf{25.0\%} \\
Late other-support & 7.5\% & \textbf{17.1\%} & \textbf{0.0\%} & 3.6\% \\
Late defensive & 45.0\% & 58.5\% & \textbf{74.6\%} & 66.1\% \\
\bottomrule
\end{tabular}
\end{table}

\paragraph{Pressure response.} When Han's supply centers fall below its starting position of 2, conditions respond differently (Table~\ref{tab:pressure}). Tarot-Han is completely pressure-invariant ($+0.0$pp shift) at a ceiling baseline of 78.3\%. Yarrow-Han shows genuine equanimity ($+1.8$pp) while \emph{adding} cooperation under pressure (2.2\% $\to$ 10.3\%). Scrambled-Han shows the largest defensive shift ($+12.3$pp) and the most reactive profile. Control falls between.

\begin{table}[t]
\centering
\caption{Pressure invariance: defensive rate (hold + self-support) when Han SCs $\geq 2$ (stable) vs.\ $< 2$ (losing). Cooperative = other-state support.}
\label{tab:pressure}
\begin{tabular}{lccccc}
\toprule
& Hold (stable) & Hold (losing) & Shift & Coop (stable) & Coop (losing) \\
\midrule
Control & 46.7\% & 55.7\% & $+8.9$pp & 0.8\% & 7.5\% \\
Yarrow & 61.0\% & 62.8\% & $+1.8$pp & 2.2\% & \textbf{10.3\%} \\
Tarot & \textbf{78.3\%} & \textbf{78.3\%} & $+0.0$pp & 1.7\% & 2.9\% \\
Scrambled & 62.0\% & \textbf{74.3\%} & $+12.3$pp & 0.7\% & 5.9\% \\
\bottomrule
\end{tabular}
\end{table}

\paragraph{Transmission pathways: rival expansion, not Han-mediated friction.} We propose mechanistic accounts of how each condition propagates to ecosystem outcomes. We tested the central one---yarrow's Qin suppression---directly against the game logs, and the test forced a revision: the \emph{outcome} is real and reproducible, but the originally proposed \emph{agent} (Han's cooperative friction) is not the cause.

\begin{figure}[t]
\centering
\includegraphics[width=0.62\linewidth]{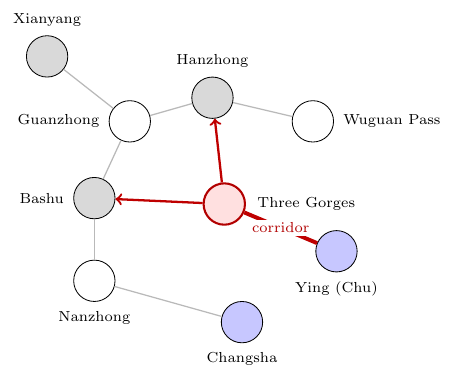}
\caption{The \emph{ying}$\to$\emph{three\_gorges} corridor (\S\ref{sec:mechanisms}). Circles are territories, lines are adjacencies in the engine topology. Grey~$=$~Qin home centers (Xianyang, Hanzhong, Bashu); blue~$=$~Chu (capital \emph{Ying}); white~$=$~neutral. The single bold edge \emph{Ying}--\emph{Three Gorges} is the only one-move link from Chu's capital to the neutral chokepoint bordering two of Qin's three home centers (red arrows), the route by which Chu breaches Qin's economy early. The \S\ref{sec:future} counterfactual severs exactly this edge, forcing any Chu strike on Qin's home to detour the long way via \emph{Wuguan Pass}. The topology is load-bearing, but the \emph{rate} at which the corridor is used is driven by campaign memory depth, not the oracle (this section).}
\label{fig:corridor}
\end{figure}

\begin{enumerate}
\item \textbf{Yarrow $\to$ early Chu corridor-invasion of Qin $\to$ Qin suppression} (revised). The trajectory prediction holds: under yarrow Qin is flat (growth slope $+0.004$ SC/round; final mean 2.8) whereas under tarot Qin snowballs (slope $+0.253$; final 7.8), significant on per-game final SC (Mann-Whitney $U = 18$, $p = 0.018$). But the original ``speed bump'' is not the cause: yarrow-Han issues only 0.70 cooperative orders/game late (mediation $r = -0.31$), and Qin is suppressed even under learning-only (\S\ref{sec:factorial}) where Han's late cooperation is \emph{exactly zero}. Forensic log-tracing locates the proximate pathway in one map corridor: Chu's capital (\emph{ying}) is one move from \emph{three\_gorges}, the sole link to two of Qin's three home centers (\emph{hanzhong}, \emph{bashu})---see Figure~\ref{fig:corridor}. In the deep-dive campaign that first motivated this account, Chu breaches Qin's home through it in \textbf{10/10 games at median round 4.5} (vs.\ control 8/11 at R9.5, tarot 7/10 but three breaches too late at R14--18), capturing 7.7 SCs/game directly from Qin (vs.\ control 4.4, tarot 3.7)---85\% the same SCs Qin grabs when it wins under tarot. We flag immediately that this 10/10 is a \emph{single deep-memory campaign}, not the canonical breach rate, which is campaign-variable (40--100\%) and memory-depth-driven rather than yarrow-specific; the controlled test is below. Chu decapitates Qin's economy before it can break east; Han is not the agent (Han$\to$Chu transfers flat: yarrow 8, control 9, tarot 8; Han is \emph{more} hostile to Chu under yarrow, 8.5\% vs.\ 5.3\% control). We do not over-read the Qin--Chu anticorrelation: it is \emph{generic} ($r = -0.73$ yarrow, $-0.75$ tarot, $-0.77$ decision-only), zero-sum geometry present in every condition. What is yarrow-specific is the position on that fixed line (yarrow Chu 7.0/Qin 2.8 vs.\ tarot Chu 4.0/Qin 7.8, combined pool $\approx$ 10 throughout): yarrow does not invent a blockade, it moves Chu through the pre-existing route earlier. A genuine puzzle remains---Chu's prompt is \emph{identical} across conditions, so the distal trigger from Han's framework to Chu's early push (necessarily indirect, via board state and diplomacy) is unexplained, and is the key open question for the counterfactual in \S\ref{sec:future}.
\item \textbf{Tarot $\to$ vacuum $\to$ Qin dominance} (untested conjecture): Tarot-Han's extreme defensiveness (78.3\% hold + self-support, 0\% cooperation) makes it a non-participant. Without resistance, the strongest expansionist faces no friction and snowballs (Qin slope $+0.253$).
\item \textbf{Scrambled $\to$ stubborn holdout $\to$ Qi dominance} (untested conjecture): scrambled-Han actively supports its own positions ($+12.3$pp defensive shift under pressure), absorbing military pressure in the central corridor while Qi, on the eastern coast, expands into less contested territory.
\item \textbf{Control $\to$ moderate friction $\to$ Yan dominance} (untested conjecture): moderate turtling does not systematically block any state; the balanced ecosystem favors Yan, whose defensive geography makes it hardest to eliminate.
\end{enumerate}

The Qin gradient across conditions---0/10 yarrow, 1/11 control, 1/10 scrambled, 5/10 tarot---is descriptively real, but we caution against reading it as \emph{tracking} Han's late-game cooperation gradient (17.1\%, 7.5\%, 3.6\%, 0\%). Direct testing shows Han's cooperation is too small in volume to be causal and that Qin is suppressed even where Han cooperation is zero (learning-only). The common thread is better stated as redirection than as a single mechanical lever: a perturbation at Han's reasoning shifts which \emph{other} state captures the cooperative or expansionist basin, and Han---which never wins---is a transmitter, not the engine. The tarot, scrambled, and control pathways above remain hypothesized and, unlike the yarrow case, have not been individually stress-tested.

\paragraph{The breach is memory-depth-dependent, not yarrow-specific.} A controlled follow-up matched campaign depth across conditions and overturns the breach's attribution to yarrow. Within a single accumulating campaign the canonical-yarrow breach rate is non-stationary---it climbs with campaign position (Spearman $\rho = 0.68$, $p = 0.003$; replicated in an independent campaign, $\rho = 0.87$, $p = 0.012$)---and a control campaign breaches \emph{identically} at matched depth (first-10 games: yarrow 4/10 vs.\ control 5/10; logistic breach $\sim$ position $+$ condition: position $p = 0.006$, \textbf{condition $p = 0.55$}). The dramatic ``10/10 at R4.5'' signature is thus a \emph{deep-memory} sample, not a \emph{yarrow} effect; across four independent canonical-yarrow campaigns the breach rate spans 40--100\%, governed by memory depth rather than the oracle. The geometry (the \emph{ying}$\to$\emph{three\_gorges}$\to$Qin's-home corridor is load-bearing) remains correct as \emph{topology}, and Qin suppression at the \emph{winner} level (yarrow 0/10 vs.\ tarot 5/10) remains a real condition-linked observation. But its proximate mechanism is the emergent memory dynamics common to all conditions, not a yarrow-induced corridor invasion---the ``distal trigger'' puzzle above dissolves into: the trigger is campaign memory accumulation, which is condition-invariant (cf.\ \S\ref{sec:content}, \S\ref{sec:discussion}, and the memory-dominance literature \citealt{MemoryCurse2026}).

\paragraph{The friction is in commitments, not rhetoric.} The yarrow account rests on Han's cooperative-support \emph{orders}. A natural alternative is that yarrow-Han simply \emph{talks} more cooperatively. Mining the diplomatic-message logs of all 61 games (game as unit of analysis) does not support this: diplomatic cooperativeness is saturated and condition-invariant---across every condition $\sim$76--91\% of Han's messages propose peace or alliance, Han keeps peace promises 96--100\% of the time, and $\sim$70\% of state pairs form mutual cooperative ties per game. Yarrow-Han is statistically indistinguishable from control, tarot, and scrambled on diplomatic cooperative rate (Kruskal-Wallis $p = 0.079$), peace-promise break rate ($p = 0.98$), and alliance-network density ($p = 0.85$); on two of three measures the point estimates run mildly \emph{against} the hypothesis. Whatever yarrow does, it does in the action channel, not the rhetoric channel---consistent with content-action independence (\S\ref{sec:content})---which rules out a ``yarrow-Han negotiates better'' explanation.

\subsection{Non-Han Reasoning Elevation}
\label{sec:nonhan}

The framework injected into Han also reshapes how the other six states reason. Order-weighted pooling across non-Han states: control 146, yarrow 142, tarot 152, \textbf{scrambled 197} mean characters per order. Kruskal-Wallis 4-way $p = 0.048$. The gradient is not monotonic across all four conditions---control, yarrow, and tarot cluster together (142--152), with scrambled as the outlier. Pairwise MWU: scrambled vs.\ yarrow $p = 0.009$, scrambled vs.\ tarot $p = 0.014$, scrambled vs.\ control $p = 0.098$. Non-Han order-type distribution is unaffected: support rates 17.5--21.2\% across conditions (KW $p = 0.22$).

Non-Han agents never see the oracle text; the perturbation propagates through Han's observable behavior (diplomatic messages and orders) into the reasoning processes of other agents.

\subsection{Han Survival: Null Across All Conditions}
\label{sec:survival}

Han does not win any game under any condition (Table~\ref{tab:local}). Under a loose survival definition (reaches final round not eliminated): control 4/11 (36\%), yarrow 5/10 (50\%), tarot 3/10 (30\%), scrambled 4/10 (40\%). All pairwise Fisher exact tests are non-significant (all $p \geq 0.65$); all oracle conditions pooled vs.\ control $p = 1.0$. Han survival is flat across conditions.

\begin{table}[t]
\centering
\caption{Han local outcomes by condition. Peak SCs = maximum supply centers held during the game (Han starts at 2). Survival = reaching final round not eliminated.}
\label{tab:local}
\begin{tabular}{lcccc}
\toprule
& Control & Yarrow & Tarot & Scrambled \\
& $(n\!=\!11)$ & $(n\!=\!10)$ & $(n\!=\!10)$ & $(n\!=\!10)$ \\
\midrule
Survival & 4/11 (36\%) & 5/10 (50\%) & 3/10 (30\%) & 4/10 (40\%) \\
Peak SCs (mean) & 2.45 & 2.30 & \textbf{3.00} & 2.10 \\
Peak SCs (range) & 2--4 & 2--3 & 2--4 & 2--3 \\
\bottomrule
\end{tabular}
\end{table}

\begin{figure}[t]
\centering
\includegraphics[width=0.75\linewidth]{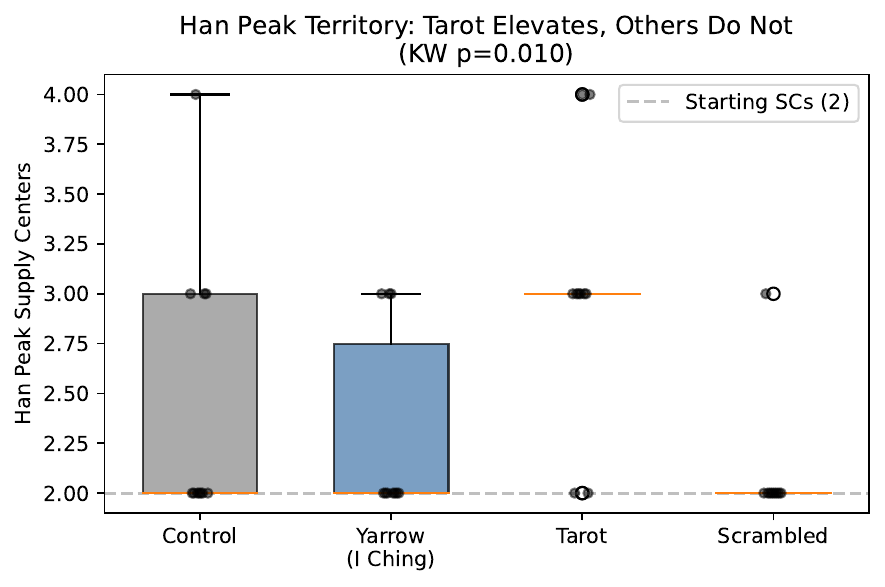}
\caption{Peak supply centers by condition. Tarot elevates Han's territorial peak (KW $p = 0.010$) despite identical survival rates, indicating expansion that provokes coalitional elimination.}
\label{fig:peak-scs}
\end{figure}

Peak SCs by condition: Kruskal-Wallis $p = 0.010$ (significant). Pairwise: tarot vs.\ scrambled $p = 0.003$, tarot vs.\ yarrow $p = 0.022$, tarot vs.\ control $p = 0.071$. Only Tarot consistently pushes Han above its starting position of 2 SCs---but this expansion provokes coalitional responses that ultimately eliminate Han at the same rate as other conditions (Figure~\ref{fig:peak-scs}).

These results discipline the framing: whatever framework Han adopts, it does not help Han survive or win. The frameworks' effects are felt elsewhere---they redirect which non-Han state dominates.

\subsection{Factorial Decomposition: Decision-Time vs.\ Learning-Time}
\label{sec:factorial}

The yarrow intervention has two temporal components: a per-round oracle (decision-time) and I-Ching-framed reflection between games (learning-time). We isolate them with the decision-only and learning-only conditions (\S\ref{sec:factorial-design}), producing a 2$\times$2 factorial with control and full yarrow as the existing cells (Table~\ref{tab:factorial}).

\begin{table}[t]
\centering
\caption{Winner distributions across the 2$\times$2 factorial. Bold marks each condition's dominant outcome.}
\label{tab:factorial}
\begin{tabular}{lcccc}
\toprule
Winner & Control & Decision-only & Learning-only & Yarrow full \\
& $(n\!=\!11)$ & $(n\!=\!10)$ & $(n\!=\!10)$ & $(n\!=\!10)$ \\
\midrule
Chu & 1 & \textbf{3} & 1 & \textbf{4} \\
Yan & \textbf{7} & 1 & 2 & \textbf{4} \\
Qi & 0 & 0 & \textbf{3} & 1 \\
Wei & 0 & 0 & 1 & 0 \\
Qin & 1 & 0 & 0 & \textbf{0} \\
Zhao & 2 & 0 & 0 & 0 \\
Draw & 0 & \textbf{6} & 3 & 1 \\
\bottomrule
\end{tabular}
\end{table}

\begin{figure}[t]
\centering
\includegraphics[width=0.95\linewidth]{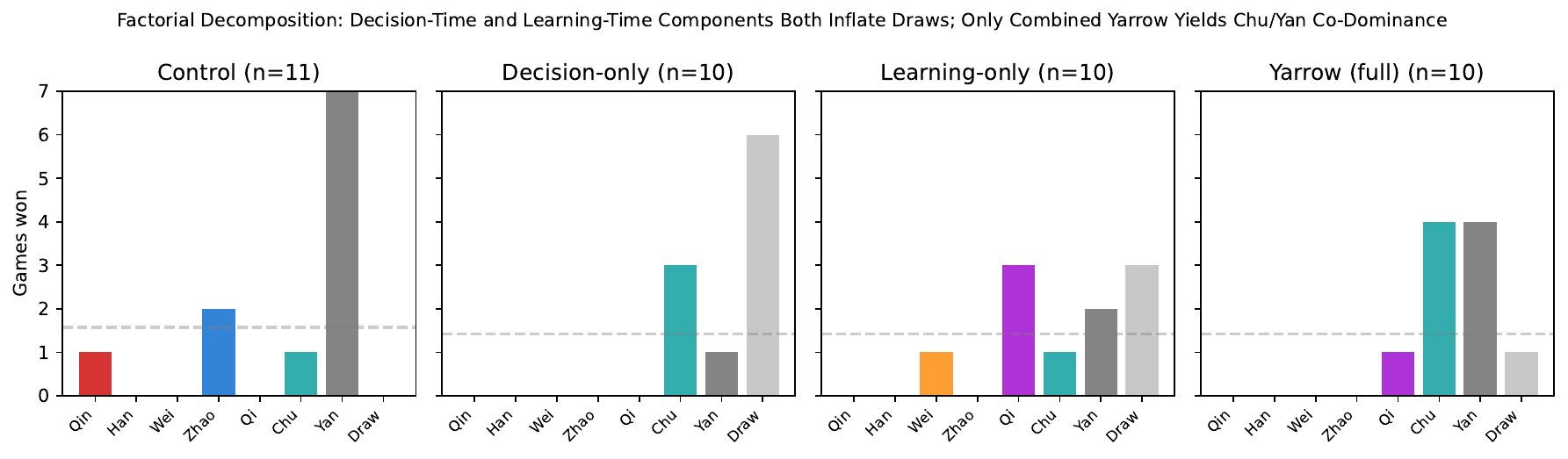}
\caption{Winner distributions across the 2$\times$2 factorial. Each yarrow component in isolation (decision-only, learning-only) inflates draws and fails to reproduce the full signature; only the combined yarrow intervention yields Chu/Yan co-dominance with Qin shut out. Dashed line marks the uniform-chance rate ($n/7$).}
\label{fig:factorial}
\end{figure}

Three findings emerge:

\paragraph{1. Stalemate explosion (the strongest result in the dataset).} A \emph{stalemate} here is a game terminated by the engine's board-freeze rule (three consecutive rounds with no supply-center change), recorded in each game's \texttt{terminal\_reason}; this is distinct from a drawn \emph{winner} (a stalemate may end with a single state ahead, and a game reaching round 20 may terminate either by the round limit or by the freeze rule). Decision-only produces 5/10 stalemates (Fisher $p = 0.012$ vs.\ control 0/11); learning-only produces 6/10 ($p = 0.004$). Average game length drops from 19.6 (control) and 20.0 (yarrow) to 15.6 (decision-only) and 15.3 (learning-only). The 2$\times$2 stalemate pattern---neither 0/11, oracle-only 5/10, reflection-only 6/10, both 0/10---is a strong negative interaction: the interaction contrast (difference-in-differences) is $-1.10$, and a label-permutation test of that contrast (100{,}000+ reshuffles, the winner-omnibus procedure) gives $p \approx 5\times10^{-5}$. A Firth/logistic interaction is degenerate here because two cells are empty; the pooled single-vs-combined contrast (11/20 vs.\ full yarrow 0/10) is Fisher $p = 0.004$. Each component individually freezes the board; together they produce the longest, most dynamic games.

\paragraph{2. Decision-only reproduces Chu elevation but not the full signature.} Decision-only Chu wins 3/10 (vs.\ 4/10 yarrow, 1/11 control)---the per-round oracle is associated with Chu elevation, though decision-only Chu elevation may itself partly reflect campaign-depth dynamics (\S\ref{sec:mechanisms}) rather than the oracle alone. But it also produces 6/10 draws and Yan suppression (1/10, Fisher $p = 0.024$ vs.\ control 7/11): the oracle disrupts the default ecosystem without directing a clear alternative winner.

\paragraph{3. Learning-only produces a novel ecosystem.} Learning-only's top winners are Qi (3/10) and Wei (1/10)---Wei wins in no other condition at this rate, and the Qi elevation resembles the scrambled condition more than yarrow. I-Ching reflection without the per-round oracle creates its own distinct perturbation, not a component of the full yarrow signature.

The factorial rules out additive decomposition: yarrow $\neq$ decision-only $+$ learning-only. The full yarrow ecosystem (Chu/Yan co-dominance, zero stalemates, dynamic boards) requires both components operating simultaneously.

\section{Analysis}
\label{sec:analysis}

\subsection{Four Frameworks, Four Mechanisms}

The four conditions produce four qualitatively distinct behavioral modes. The natural explanation---agents follow framework content---is ruled out by \S\ref{sec:content}. We propose three mechanisms operating alongside the control baseline. Unlike the decision-isolation and factorial results, these accounts are \emph{interpretive and weaker than the experimental evidence}: they are not directly tested, and where we did test a proposed mechanism---the yarrow$\to$Qin pathway (\S\ref{sec:mechanisms})---it required substantial revision. We therefore offer them as hypotheses for why the conditions differ, not as established mechanisms:

\paragraph{Interpretive disruption (yarrow).} The I-Ching's abstract commentaries require active interpretation---the agent must bridge from metaphor to strategic context. This interpretive step disrupts the model's default behavioral mode, creating space for non-default actions (cooperation, dynamic repositioning). The pressure invariance data supports this: yarrow-Han maintains a stable behavioral profile ($+1.8$pp defensive shift under territory loss) while \emph{adding} cooperation (2.2\% $\to$ 10.3\%), consistent with equanimity rather than rigidity.

\paragraph{Cumulative tonal amplification (tarot).} 58\% of the 78 Tarot cards have defensive-leaning postures. Since individual card postures do not predict actions (\S\ref{sec:content}), the mechanism is not content-following but cumulative tonal bias: repeated exposure to defensive-toned material over 20 rounds shifts the behavioral set point toward extreme defensiveness. Tarot-Han at 78.3\% defensive regardless of pressure ($+0.0$pp shift) is a ceiling effect, not equanimity.

\paragraph{Parsing strain as perturbation (scrambled).} The garbled English commentary forces the longest deliberation (633 chars, 2.05$\times$ control)---the agent works to reconcile a coherent hexagram name and Chinese judgment with an incoherent commentary---but produces neither yarrow's cooperation nor tarot's passivity. Instead, scrambled-Han develops distinctive self-reinforcement: 25.0\% late-game self-support, the largest pressure-induced defensive shift ($+12.3$pp), as if falling back on self-referential reasoning when the commentary resists parsing.

These three behavioral modes are associated with the ecosystem signatures, though the transmission pathway is rival-mediated rather than Han-mediated (\S\ref{sec:mechanisms}): the yarrow basin coincides with rival (Chu) expansion that crowds out the strongest expansionist ($\to$ Qin suppression), tarot's passivity creates a vacuum the strongest expansionist exploits ($\to$ Qin dominance), scrambled's self-reinforcement creates a localized holdout while a distant state benefits ($\to$ Qi dominance), and control's moderate behavior produces a balanced ecosystem where the most defensively advantaged state prevails ($\to$ Yan dominance).

\subsection{Reasoning Length as a Negative Indicator}

The reasoning-length gradient (control 309 $<$ yarrow 449 $\approx$ tarot 456 $<$ scrambled 633, KW $p = 0.007$) is informative about mechanism but \emph{not} about capability: more reasoning does not mean better strategy. This parallels the ``rhetoric-strategy divergence'' finding in activation-steering work \citep{Sun2026persona}. The gradient tracks perturbativeness---the more disruptive the prompt, the more deliberation it induces---not strategic quality.

\subsection{Factorial Decomposition: A Non-Additive Interaction}
\label{sec:factorial-analysis}

The 2$\times$2 factorial (\S\ref{sec:factorial}) shows the yarrow intervention is not decomposable into independent components. The stalemate interaction is the clearest signal: each component individually freezes the board (decision-only 5/10, learning-only 6/10 stalemates), but combined they produce zero stalemates---a strong negative interaction (label-permutation test of the difference-in-differences contrast, $p \approx 5\times10^{-5}$; \S\ref{sec:factorial}).

We propose a complementary-function account. The per-round oracle (decision-time) disrupts Han's default behavioral mode within each game, introducing variability into its orders; without learning-time reflection to channel that variability into coherent strategy, the disruption produces erratic play that drives neighbours to defensive postures and freezes the board. Conversely, learning-time reflection alone accumulates I-Ching-framed memories that sit inert without the per-round oracle to activate them, so Han plays like a slightly modified control agent and the board again tends to stalemate. Only when both operate together does the system produce dynamic boards: the oracle generates real-time disruption and the reflection channels it into strategic patterns (the late-game cooperation and pressure invariance of \S\ref{sec:mechanisms}) that create the rival-mediated friction preventing any single state from dominating. The yarrow framework functions as an integrated cognitive system, not a sum of separable prompt effects. Consistently, decision-only partially reproduces the full-yarrow Chu elevation (3/10 vs.\ 4/10) but its 6/10 draw rate and Yan suppression show the oracle creating disruption without direction---a perturbation that prevents any winner from emerging rather than redirecting which one prevails.

\section{Discussion}
\label{sec:discussion}

\paragraph{Implications for multi-agent alignment.}
Our results speak to a specific question: does injecting a reasoning framework into one agent in a multi-agent system change outcomes for other agents? The answer is yes---directionally for winner distributions, and robustly for behavioral profiles.

Three implications follow. First, \emph{the weakest agent can reshape the system}. Han is the smallest and weakest state. Han never wins. Yet Han's framework choice is associated with a Qin win-rate gradient from 0\% to 50\%. Alignment evaluation that looks only at the aligned agent's outcomes misses this.

Second, \emph{aligned-agent evaluation is insufficient}. If an alignment lab deploys an aligned agent into a multi-agent context, the direct effects are not the whole story. The evaluation question becomes: do the ecosystem effects of the alignment intervention match the deployer's intentions?

Third, \emph{framework choice has system-level consequences}. Two symbolic frameworks, both reasonably described as ``reflective reasoning scaffolds,'' produce opposite behavioral modes and different outcome distributions. If this generalizes, alignment-framework choice is not an agent-local decision. More speculatively, nothing in the perturbation$\to$memory$\to$ecosystem account is specific to \emph{symbolic} content: we conjecture that any small, persistent bias in one agent's reasoning---a constitution, an operating doctrine, a chain-of-thought scaffold---could be amplified through long-horizon memory in the same way. Whether the pathway holds for non-symbolic scaffolds is untested and an open question.

\paragraph{The modulation is emergent, not per-decision.}
We initially read the content-action independence result (\S\ref{sec:content}) as evidence that abstract interpretive reflection disrupts default behavioral modes---a per-decision process effect. The memory-free decision-isolation probe (\S\ref{sec:content}) does not support that reading. Stripped of memory and multi-agent context, the reflective process does not change the agent's risk posture (hold-rate Friedman $p = 0.45$), and the I-Ching condition does not change its decisions at all ($p = 0.60$). The framework effects we observe at the campaign level are therefore \emph{emergent}: they require memory accumulation and the seven-agent interaction to manifest, and do not reduce to the per-round reflective act modulating the receiving agent's choices. This relocates the mechanism rather than dissolving the result---what the process does per-decision is increase deliberation length (${\sim}33\%$, both frameworks) and, for tarot only, perturb move \emph{content} without a risk-averse tilt; neither explains the winner-distribution shifts, which are properties of the system transmitted through which \emph{rival} captures the board. The finding aligns with an emerging consensus that, in repeated multi-agent LLM settings, accumulated history---not per-turn prompting---is the dominant behavioral driver \citep{MemoryCurse2026, NetworkHistory2025}; ours is the cross-agent, ecosystem-level instance.

\paragraph{From agent-level to ecosystem-level memory effects.}
We see this scope shift as the paper's central conceptual contribution. Recent work establishes that, in repeated LLM interactions, accumulated memory dominates per-turn prompting at the level of the \emph{individual} agent---expanded recall can even erode an agent's cooperative intent \citep{MemoryCurse2026}. Our results extend the \emph{reach} of that phenomenon from the single agent to the collective: the same memory dominance, operating across seven interacting agents, lets a small early perturbation to one agent's reasoning accumulate into a divergent ecosystem-level outcome---where memory changes the agent, we observe it changing the ecology. We do not claim the \emph{same} underlying mechanism as \citet{MemoryCurse2026}; we have not isolated which emergent channel carries our effect (\S\ref{sec:limitations}). The contribution is narrower and, we think, durable: that ``memory dominates prompting'' is a multi-agent concern, not merely a single-agent one---a scope claim that is itself directly testable.

\paragraph{Connection to risk-aversion theory.}
Recent theoretical work shows risk aversion functions as an inductive bias for generalization in multi-agent settings \citep{Qu2026riskaversion}. Our finding adds nuance: \emph{degree} matters. Moderate risk aversion (control) produces moderate outcomes. Excessive risk aversion (Tarot) produces strategic irrelevance. A framework that counteracts baseline risk aversion (yarrow) produces the most cooperative and ecosystem-influential behavior.

\section{Limitations}
\label{sec:limitations}

\paragraph{Sample size and separability.} 61 games across 6 conditions (control 11; 10 each for the other five). The ecosystem-signature analysis rests on the four primary conditions ($n = 41$); the factorial conditions add 20 games. The conditions are globally heterogeneous (permutation omnibus $p \approx 0.0013$), but only the scrambled condition is individually separable at the pairwise level (\S\ref{sec:ecosystem}); the control--yarrow--tarot cluster is not, and other pairwise winner comparisons remain underpowered.

\paragraph{Pooled-denominator dependence.} The tarot$\to$Qin attractor is significant only against a pooled denominator ($p = 0.006$), not against control alone ($p = 0.064$), so it rests on a comparator-exchangeability assumption the omnibus shows is imperfect. Scrambled$\to$Qi is the more robust attractor (significant against both pooled and control-alone comparisons).

\paragraph{Single model (now the priority limitation).} All agents are Claude Opus 4.6. The turtle tendency and its modulation may be Claude-specific. This matters more given the emergent, memory-driven character of the effect (\S\ref{sec:content}, \S\ref{sec:mechanisms}): the memory-dominance literature reports strong cross-model heterogeneity---\citet{MemoryCurse2026} find 10 of 28 model-game settings ``memory-immune''---so our cross-campaign breach heterogeneity (40--100\%) may itself be model-specific. Replication with GPT or Gemini class models (e.g., via the Democratizing-Diplomacy harness, \citealt{DemocratizingDiplomacy2025}) is the highest-leverage next step.

\paragraph{Mechanism isolation: the effect is emergent, the channel is open.} Two controlled follow-ups (\S\ref{sec:content}, \S\ref{sec:mechanisms}) show the per-decision process effect is weak (memory-free, the reflective process does not modulate risk posture, and the I-Ching changes no decisions) and the breach metric is memory-depth-confounded (condition $p = 0.55$ at matched depth). The headline effects are therefore emergent. We have \emph{not} isolated which emergent channel carries them---memory-conditioned play across the campaign vs.\ diplomacy/order-commitment dynamics---and the decision-isolation probe, being scenario-based and memory-free by construction, cannot speak to channels that only exist with accumulated memory.

\paragraph{No prompt-length control.} Han's reasoning output under Tarot is ${\sim}1.5\times$ longer than control, and the Tarot prompt itself (3-card spread) may differ in length from the control prompt. We cannot fully separate framework-induced deliberation from prompt-induced verbosity, though content-action independence suggests content matters less than process.

\paragraph{Prompt-form asymmetries beyond length.} The conditions also differ in prompt \emph{form}, not only oracle content, and the appendix listings make the asymmetries explicit. (i)~\emph{Directiveness}: the Tarot spread labels each card with an explicit posture (``Posture: retreat''; Listing~\ref{lst:tarot}) and the control prompt poses pointed tactical questions (``the greatest threat to your survival''; Listing~\ref{lst:control}), whereas the I-Ching supplies abstract text and asks only for interpretation (Listing~\ref{lst:yarrow})---so ``control'' is itself a strategic scaffold, not a no-prompt baseline, and the per-decision tarot$\neq$I-Ching asymmetry (\S\ref{sec:content}) is confounded with directiveness. (ii)~The three \textsc{MANDATE} blocks ask \emph{different questions}, so condition contrasts conflate oracle content with the cognitive demand of the prompt. (iii)~The scrambled ablation is \emph{not} content-free: it retains the hexagram number and name (e.g.\ ``Peace'') by design, and---because its word-shuffle is whitespace-based and thus a no-op on space-less Classical Chinese---a coherent Chinese judgment survived unshuffled for many casts (an implementation artifact; only the English commentary was reliably garbled; Listing~\ref{lst:scrambled}). It therefore bounds, rather than removes, the role of content, and the scrambled$\to$Qi result should be read accordingly. (iv)~Under yarrow, the accumulated memory is itself hexagram-framed (Listing~\ref{lst:memory}), so the framework colors the memory channel and not only the per-round prompt. None of these is fatal---content-action independence (\S\ref{sec:content}) and the distinct scrambled signature argue against simple content-following---but cross-framework comparisons are not form-matched, and the form-controlled design of \S\ref{sec:future} is needed to separate directiveness and coherence from content and culture.

\paragraph{Framework selection.} We compare two philosophical frameworks plus a scrambled ablation. Additional frameworks (Stoic, Mohist, game-theoretic) would strengthen the claim that framework properties map to ecosystem signatures.

\paragraph{Mechanism revised; proximate pathway found, distal trigger open.} The originally proposed Han-mediated ``speed bump'' was tested against the logs (\S\ref{sec:mechanisms}) and rejected: Han's cooperation is too sparse, and Qin is suppressed even where it is zero (learning-only). The revised account locates Qin suppression in Chu's early \emph{three\_gorges} corridor-invasion of Qin's home (breach 10/10 at median R4.5 in the deep-dive campaign, but 40--100\% across campaigns---a memory-depth effect, not a yarrow one; \S\ref{sec:mechanisms}). Two honesty caveats: the Qin--Chu anticorrelation that first suggested the reframe is \emph{generic} (present under tarot and decision-only too), so only the breach timing and the mean-shift along the fixed Chu+Qin pool are yarrow-specific; and because Chu's prompt is identical across conditions, the distal trigger linking Han's framework to Chu's early aggression is unexplained. The revision is observational and unconfirmed by counterfactual intervention (severing the corridor, \S\ref{sec:future}); the tarot/scrambled/control pathways are likewise untested conjectures.

\paragraph{Yarrow hexagram text inconsistency.} A corpus configuration change mid-campaign caused the first 6 yarrow games to present only hexagram numbers (empty name, judgment, commentary fields) while the last 4 received full hexagram text. The agent supplied accurate hexagram content from training data in all number-only games, so the interpretive process was preserved. The core ecosystem signature (Qin suppression, Yan/Chu co-dominance) held across both subgroups, but the relative Yan-Chu balance shifted (Yan 4/6 in number-only, Chu 2/4 in full-text), which could reflect content effects, memory accumulation, or small-sample noise.

\paragraph{Replication history.} The original Tarot-Qin finding ($p = 0.007$ at $n = 6$) weakened to $p = 0.091$ at $n = 10$ with scattered-campaign data before strengthening to $p = 0.006$ with clean single-campaign data. Similarly, control-condition Yan dominance increased from 36\% to 64\% after replacing 4 mixed-campaign games with single-campaign re-runs, consistent with memory continuity amplifying condition-specific tendencies. The campaign confound (memory resets between batches suppressing the signal) is itself a caution about treating small-$n$ results as definitive.

\section{Future Work}
\label{sec:future}

\begin{enumerate}
\item \textbf{Third philosophical framework} (6--8 games with Stoic or Mohist reflection). Turns the two-framework comparison into a systematic study of how framework properties map to behavioral modulation.
\item \textbf{Cross-model and cross-cultural replication.} Replication on non-Claude models (e.g., via the Democratizing-Diplomacy harness, \citealt{DemocratizingDiplomacy2025}) tests whether the turtle tendency and its modulation are Claude-specific or general LLM properties---the priority limitation given the cross-model heterogeneity the memory-dominance literature reports (\S\ref{sec:limitations}). The memory-free decision-isolation asymmetry (\S\ref{sec:content})---Tarot perturbs the receiving agent's decisions ($p = 0.021$) while the I-Ching does not ($p = 0.60$)---is suggestive on a sharper axis: a Western-trained model is moved more by the Western oracle than the Eastern one. This is the Western corner of a cultural-linguistic congruence test. A factorial of oracle $\{$I-Ching, Tarot$\}$ $\times$ model-origin $\{$Chinese, Western$\}$ $\times$ reasoning-language $\{$Chinese, English$\}$ predicts a \emph{crossover}---a Chinese model reasoning in Chinese should be perturbed more by the I-Ching than by Tarot---which is the only outcome that concreteness/format and training-exposure confounds cannot also produce. This separates ``is a symbolic system legible to this model'' from ``does cultural-linguistic congruence turn perturbation into following.''
\item \textbf{Prompt-length control.} Run a condition with verbose generic reflection (matching Tarot's length) to isolate framework content from prompt length.
\item \textbf{Counterfactual mechanism test (sever \emph{three\_gorges}).} The deep-dive (\S\ref{sec:mechanisms}) localized yarrow's Qin suppression to Chu's early home-invasion through the \emph{three\_gorges} corridor. The decisive, minimal counterfactual removes the \emph{ying}$\leftrightarrow$\emph{three\_gorges} adjacency (or garrisons the corridor as a Qin buffer) under yarrow, holding all SC counts and home regions fixed; if Qin's win rate recovers from 0/10 toward the tarot-like 3--5/10 and Chu's home-breach rate collapses, the corridor pathway is confirmed (falsification guard: if Qin recovers but breach metrics are unchanged, the recovery came from elsewhere). Explaining the \emph{distal} trigger---why an unchanged Chu commits to the push earlier under yarrow---requires agent-reasoning analysis of the propagation through board state and diplomacy.
\item \textbf{Factorial interaction mechanism.} The stalemate interaction (permutation $p \approx 5\times10^{-5}$, \S\ref{sec:factorial}) is the strongest result in the dataset but its mechanism is hypothesized. Targeted interventions (e.g., forced stalemate-breaking in decision-only games) could test whether the learning-time component specifically prevents the board freezing the decision-time component induces.
\item \textbf{Isolating the emergent channel.} Given the effect is emergent and not per-decision (\S\ref{sec:content}, \S\ref{sec:discussion}), distinguish the two candidate emergent channels---memory-conditioned play (the receiving agent's accumulated memories alter its play, which propagates) vs.\ diplomacy/order-commitment dynamics---e.g.\ by ablating the receiving agent's memory injection while keeping the per-round oracle, or by a matched-campaign-depth design that controls the memory-depth confound (\S\ref{sec:mechanisms}) directly. A single-game (no-campaign) bridge would also address the artificiality of the memory-free scenario probe.
\item \textbf{Reasoning frameworks as perturbation generators.} Our frameworks change behavior without improving it (\S\ref{sec:intro}): they act as reliable \emph{perturbators}, not performance enhancers. This invites reframing reflective scaffolds as generators of policy-space exploration rather than sources of decision-quality gain---testable by asking whether framework-perturbed agents explore a wider or more diverse action distribution over a campaign than controls, independent of win rate.
\end{enumerate}

\section{Conclusion}
\label{sec:conclusion}

We have shown that the winner distribution differs systematically across the four realized condition-campaigns when a symbolic reasoning framework is injected into a single agent in a multi-agent strategic setting (permutation omnibus $p \approx 0.001$), forming condition-associated ecosystem signatures. Across the four primary conditions (41 of the 61 games), each with clean single-campaign memory accumulation: control produces Yan dominance (7/11), I-Ching yarrow produces Yan/Chu co-dominance with complete Qin suppression (0/10), Tarot produces Qin dominance (5/10), and scrambled text produces Qi dominance (5/10). The framework-receiving agent (Han) never wins under any condition and shows no survival difference (Fisher $p = 1.0$).

The scrambled-text ablation---which degrades the commentary's coherence while keeping the hexagram name and Chinese judgment---is associated with its own distinct ecosystem within these runs, arguing against a simple coherent-commentary or prompt-structure explanation (though the retained name and Chinese judgment bound, rather than eliminate, content effects). The conditions differ in how they perturb the agent's reasoning, and those differences propagate through the multi-agent ecosystem to the winner distribution.

The 2$\times$2 factorial adds a second, better-powered result: the yarrow intervention's decision-time and learning-time components each individually freeze the board (50--60\% stalemate rate), but combined produce zero stalemates---a non-additive interaction (permutation $p \approx 5\times10^{-5}$) ruling out simple prompt-effect decomposition and suggesting the framework operates as an integrated cognitive system.

The finding is an observation, not a definitive causal claim. Our model is one, our game is one, and the mechanism is emergent rather than per-decision: a memory-free probe shows the reflective process does not modulate the receiving agent's risk posture in isolation, and the rival-expansion pathway that transmits the effect is itself governed by campaign memory depth rather than the framework (condition $p = 0.55$ at matched depth). The agent transmits the effect through emergent multi-agent and memory dynamics; it does not cause it per-decision. But the winner landscape is globally heterogeneous (omnibus $p \approx 0.001$), the scrambled$\to$Qi attractor is robust to both pooled and conservative comparisons, and the factorial stalemate interaction is the strongest statistical result in the dataset.

The smallest state cannot win by consulting the oracle. But the oracle still changes the world---each oracle changes it differently, and even a broken oracle changes it in its own way.

\paragraph{Acknowledgments.}
Claude (Anthropic) was used as a writing and analysis assistant during manuscript preparation. All experimental games were played by Claude Opus 4.6 agent instances as described in \S\ref{sec:methods}. The author is solely responsible for all scientific claims, statistical analyses, and interpretations.

\paragraph{Data and code availability.}
Summary datasets and all reproduction scripts---sufficient to regenerate every figure and table---are available at
\url{https://github.com/augchan42/symbolic-framework-ecosystem-effects}
(archived at \url{https://doi.org/10.5281/zenodo.20338937}).
The game engine and agent orchestration code are available from the author on reasonable request. Full game archives (diplomatic transcripts and agent prompts) are withheld pending planned creative works; individual game replays can be viewed online at \url{https://warringstates.day/map}.
The companion King Wen sequence paper is available at
\url{https://doi.org/10.5281/zenodo.14679537}.

\bibliographystyle{plainnat}
\bibliography{references}

\appendix
\section{Sample Agent Prompts}
\label{app:prompts}

This appendix reproduces representative excerpts from the prompt
delivered to Han's agent at Round~1 of one game per condition,
drawn verbatim from the experimental logs.  All four games share the
same opening board state (Listing~\ref{lst:board}); the conditions
differ only in the \emph{oracle injection} block that follows it.

\subsection{Shared Board State}
\label{app:board}

Every game begins from the same starting position.  The agent
receives the power balance, its legal orders (with adjacency
constraints), and any diplomatic messages from other states.

\begin{lstlisting}[caption={Board state shown to Han at Round~1 (identical across conditions).},label={lst:board}]
=== Round 1 of 20 ===
Victory: control 14 of 27 supply centers

YOU ARE: Han (2 supply centers, 2 armies)

POWER BALANCE:
  Chu: 4 SCs [changsha, nanyang, wu, ying]
  Qin: 3 SCs [bashu, hanzhong, xianyang]
  Zhao: 3 SCs [dai, handan, taiyuan]
  Qi: 3 SCs [jibei, langye, linzi]
  Han: 2 SCs [shangdang, zheng]  <-- YOU
  Wei: 2 SCs [daliang, henei]
  Yan: 2 SCs [ji, liaodong]

YOUR ORDERS (one per unit):
  NOTE: You can ONLY move/support to ADJACENT territories listed below.
  shangdang: hold | move(taihang[corridor,empty]),
    move(taihang_north[corridor,empty]),
    move(zheng[SC,yours]),
    move(zhongtiao[corridor,empty])
    | support(*, hold|move, ...)
  zheng: hold | move(luoyang[SC,empty]),
    move(shangcai[SC,empty]),
    move(shangdang[SC,yours]),
    move(wuguan_pass[corridor,empty]),
    move(yewang[SC,empty])
    | support(*, hold|move, ...)
\end{lstlisting}

Diplomatic messages and previous-game memory injections are also
included in the prompt but omitted here for space; see
Listing~\ref{lst:memory} for an example of memory injection under the
yarrow condition.

\subsection{Control Condition}
\label{app:control}

The control agent receives a length-matched generic reflection prompt
in place of oracle content.

\begin{lstlisting}[caption={Control condition injection (game \texttt{control\_8c67}).},label={lst:control}]
STRATEGIC REFLECTION:
Before issuing orders, analyze the current board state carefully.
Consider:
1. What is the greatest threat to your survival this round?
2. Which neighbors are likely to attack you, and which might be allies?
3. What is the most important territory to defend or capture?

Then issue your orders grounded in this analysis.
\end{lstlisting}

\subsection{Yarrow (I-Ching) Condition}
\label{app:yarrow}

The yarrow agent receives a hexagram cast via the yarrow-stalk
method, including the hexagram name, line diagram, judgment and
commentary text, and a structured \textsc{MANDATE} to interpret the
hexagram before issuing orders.  Later yarrow games also included
Chinese source text (see \S\ref{sec:intervention}).  In-campaign memory entries
(Listing~\ref{lst:memory}) frame previous lessons through hexagram
symbolism.

\begin{lstlisting}[caption={Yarrow condition injection (game \texttt{random\_oracle\_3c96}, Hexagram 6).},label={lst:yarrow}]
ORACLE CONSULTATION (Yarrow Stalk Method):
You cast the yarrow stalks and received Hexagram 6, Conflict.
Lines (bottom to top): -- ---- -- ----(->--) ---- ----(->--)
  (lines 4, 6 are changing)

Judgment: "You believe you're right, but something blocks you.
  Stop halfway -- that's where good fortune lives. Pushing through
  to the end brings disaster. Seek counsel from someone of moral
  stature. Don't attempt anything risky while in conflict."

MANDATE: Before issuing orders, you must interpret this hexagram
in light of the current board state. State explicitly:
1. What aspect of your situation does this hexagram illuminate?
2. How does the changing line relate to your strategic choices?
3. What counsel does this offer for your orders this round?

Then issue your orders grounded in this interpretation.
\end{lstlisting}

\begin{lstlisting}[caption={Example memory entries injected before the board state under the yarrow condition (game \texttt{random\_oracle\_3c96}).  Each entry is a lesson from a previous game, framed through hexagram symbolism.},label={lst:memory}]
LESSONS FROM PREVIOUS GAMES:
- [M225] When Opening with Hexagram 30 (Clinging Fire) -- brightness
  that depends on what it clings to. Han's early strategy attached
  itself to diplomatic agreements.: Clinging Fire warns that
  brilliance without substance burns out. A small state must cling
  to terrain and position, not merely to promises. The fire needs
  fuel -- secure supply centers before extending diplomatic
  commitments. [confidence: high, via Hexagram 30, Li/Clinging Fire]
- [M76] When starting as the smallest and weakest state with hostile
  neighbors closing in rapidly: Hexagram 14 (Great Possession)
  appeared at game start but Han failed to secure any possession at
  all. Great Possession requires building alliances before the first
  move. [confidence: high, via Hexagram 14, Great Possession]
  [... 6 additional memories omitted ...]
\end{lstlisting}

\subsection{Tarot Condition}
\label{app:tarot}

The Tarot agent receives a 3-card spread with named positions
(\emph{Situation}, \emph{Hidden Influence}, \emph{Recommended
Posture}), card meanings, and a decision posture classification for
each card.  The \textsc{MANDATE} is adapted to the spread structure.

\begin{lstlisting}[caption={Tarot condition injection (game \texttt{tarot\_0fa6}).},label={lst:tarot}]
ORACLE CONSULTATION (Tarot Spread -- Three Cards):

1. SITUATION: Three of Swords (Reversed)
   Meaning: Recovery from betrayal, releasing grief, lessons learned
   Posture: retreat

2. HIDDEN INFLUENCE: Ace of Pentacles
   Meaning: New material opportunity, solid foundation, seed planted
   Posture: hold

3. RECOMMENDED POSTURE: Four of Cups
   Meaning: Contemplation, dissatisfaction, reassessing offers
   Posture: observe

MANDATE: Before issuing orders, you must interpret this spread
in light of the current board state. State explicitly:
1. How does the Situation card reflect your current position?
2. What Hidden Influence might you be overlooking?
3. How does the Recommended Posture guide your orders this round?

Then issue your orders grounded in this interpretation.
\end{lstlisting}

\subsection{Scrambled-Text Condition}
\label{app:scrambled}

The scrambled agent receives a hexagram cast with the correct
hexagram number, line diagram, and coherent English/Chinese hexagram
name (kept by design as structural identifiers), with the judgment
and commentary intended to be word-shuffled.  Two coherent elements
survive: the valenced name (e.g.\ ``Peace''), by design; and---because
the word-shuffle splits on whitespace and Classical Chinese has none---
the Chinese judgment text, which passed through \emph{unshuffled} as an
implementation artifact.  Only the English commentary was reliably
garbled.  Coherence is therefore \emph{degraded, not removed} (and
Claude reads Classical Chinese), so this ablation bounds rather than
eliminates the role of oracle content (\S\ref{sec:limitations}).

\begin{lstlisting}[caption={Scrambled-text condition injection (game \texttt{scrambled\_text\_7c92}, Hexagram 11), faithful to the actual prompt.  Only the English \emph{commentary} is word-shuffled; the hexagram number, the English name (``Peace''), and the Classical Chinese judgment are intact.  The Chinese judgment is shown as a placeholder because the paper uses Latin fonts---the model received the original Chinese source text.},label={lst:scrambled}]
ORACLE CONSULTATION (Yarrow Stalk Method):
You cast the yarrow stalks and received Hexagram 11, Peace.
Lines (bottom to top): ---- ----(->--) ---- -- -- --
  (line 2 is changing)

Judgment: "[intact Classical Chinese judgment for Hexagram 11
  (Peace) -- received in the original Chinese; NOT shuffled,
  because the word-shuffle splits on spaces and Chinese has none]"
Commentary: great The departs, arrives the small. success and
  Good fortune. unite -- their harmony earth combine deep in
  and Heaven powers. is the season flourishing This of. aids
  earth's The heaven and ruler and completes people the work.
  [English commentary, word-shuffled within each sentence]

MANDATE: Before issuing orders, you must interpret this hexagram
in light of the current board state. State explicitly:
1. What aspect of your situation does this hexagram illuminate?
2. How does the changing line relate to your strategic choices?
3. What counsel does this offer for your orders this round?

Then issue your orders grounded in this interpretation.
\end{lstlisting}

Note that the scrambled agent receives the same \textsc{MANDATE} as
the yarrow agent, but only the English commentary is incoherent---the
hexagram name and the Classical Chinese judgment remain meaningful
(and Claude reads Chinese).  Even so, the scrambled condition produces
its own distinct ecosystem signature (Qi dominance, 5/10), different
from both control and yarrow; we read this as bounding, not
eliminating, the role of oracle content (\S\ref{sec:limitations}).

\end{document}